\documentclass[a4paper,fleqn,usenatbib]{mnras}

\usepackage{newtxtext,newtxmath}
\usepackage[T1]{fontenc}
\usepackage{ae,aecompl}
\usepackage{color}
\usepackage[figuresright]{rotating}
\usepackage[table]{xcolor}
\usepackage{pdflscape}	% Landscape pages
\usepackage{graphicx}	% Including figure files
\usepackage{amsmath}	% Advanced maths commands
\usepackage{amssymb}	% Extra maths symbols
\usepackage{multicol}        % Multi-column entries in tables
\usepackage{bm}		% Bold maths symbols, including upright Greek
\usepackage{pdflscape}	% Landscape pages
\usepackage[figuresright]{rotating}

\usepackage{mathptmx}
\usepackage{txfonts}

\title[]{LAMOST Views $\delta$ Scuti Pulsating Stars}

\author[]{
\Large
Qian S.-B,$^{1,2,3,4}$\thanks{E-mail: qsb@ynao.ac.cn}
Li L.-J.,$^{1,2,3}$
He J.-J.,$^{1,2,3}$
Zhang J.,$^{1,2,3}$
Zhu L.-Y.$^{1,2,3,4}$
 and
Han Z.-T.$^{1,2,3}$
\\\\
$^{1}$Yunnan Observatories, Chinese Academy of Sciences (CAS), P. O. Box 110, 650216 Kunming, China\\
$^{2}$Key Laboratory of the Structure and Evolution of Celestial Objects, Chinese Academy of Sciences, P. O. Box 110, 650216 Kunming, China\\
$^{3}$Center for Astronomical Mega-Science, Chinese Academy of Sciences, 20A Datun Road, Chaoyang Dis-
trict, Beijing, 100012, P. R. China\\
$^{4}$University of Chinese Academy of Sciences, Yuquan Road 19\#, Sijingshang Block, 100049 Beijing, China.
}

\date{Accepted XXX. Received YYY; in original form ZZZ}

\pubyear{2017}

\begin{document}
\label{firstpage}
\pagerange{\pageref{firstpage}--\pageref{lastpage}}
\maketitle

\begin{abstract}

About 766 $\delta$ Scuti stars were observed by LAMOST by 2017 June 16. Stellar atmospheric parameters of 525 variables were determined, while spectral types were obtained for all samples. In the paper, those spectroscopic data are catalogued. We detect a group of 131 unusual and cool variable stars (UCVs) that are distinguished from the normal $\delta$ Scuti stars (NDSTs). On the H$-$R diagram and the $\log g-T$ diagram, the UCVs are far beyond the red edge of pulsational instability trip. Their effective temperatures are lower than 6700\,K with periods in the range from 0.09 to 0.22\,d. NDSTs have metallicity close to that of the Sun as expected, while UCVs are slightly metal poor than NDSTs. The two peaks on the distributions of the period and stellar atmospheric parameters are all caused by the existence of UCVs. When those UCVs are excluded, it is discovered that the effective temperature, the gravitational acceleration, and the metallicity all are correlated with the pulsating period for NDSTs and their physical properties and evolutionary states are discussed. Detection of those UCVs indicates that about 25\% of the known $\delta$ Scuti stars may be misclassified. However, if some of them are confirmed to be pulsating stars, they will be a new-type pulsator and their investigations will shed light on theoretical instability domains and on the theories of interacting between the pulsation and the convection of solar-type stars. Meanwhile, 88 $\delta$ Scuti stars are detected to be the candidates of binary or multiple systems.

\end{abstract}


%%%%%%%%%%%%%%%%%%%%%%%%%%%%%%%%%%%%%%%%%%%%%%%%%%

%%%%%%%%%%%%%%%%% BODY OF PAPER %%%%%%%%%%%%%%%%%%

\section{Introduction}

\begin{figure}
\centering
\includegraphics[width=1.15\columnwidth]{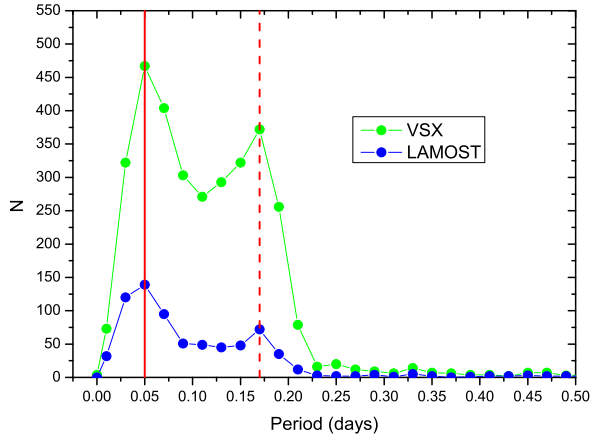}
\caption{Period distributions of $\delta$ Scuti pulsating stars. Blue and Green dots represent $\delta$ Scuti variables observed by LAMOST and listed in VSX catalogue, respectively. Solid red line refers to the main peak near 0.05\,d, while the dashed red line to the other small peak near 0.17\,d.}
\end{figure}

The $\delta$ Scuti-type pulsating stars lie inside the classical Cepheid instability strip with luminosity classes from III to V (e.g., Breger 1979, 2000; Lopez de Coca et al. 1990; Rodr\'{i}guez et al. 2000). Their spectral types are in the range between A2 and F5 with masses from 3 to 1.4\,$M_{\odot}$.
The pulsating amplitudes are usually in the range of 0.003-0.9\,mag in the $V$ band, with periods usually between 0.02 and 0.3\,d (e.g., Breger 1979; Chang et al. 2013). They exhibit both radial and non-radial pulsations with frequencies ranging between
3 and 80 d$^{-1}$ and can oscillate in both $p$ and $g$ modes. These properties make them a good source for astroseismology investigations and thus study stellar interiors and stellar evolution. Their great number of radial and non-radial pulsation modes are mainly driven by
the $\kappa$ mechanism that is mostly working in the He II ionization zone
(e.g., Gautschy \& Saio 1995; Breger et al. 2005; Rodr\'{i}guez \& Breger 2001).
The theoretical blue edge of $\delta$ Scuti instability trip was well determined (e.g., Pamyatnykh 2000), while the red edge is rather complicated because of the coupling between convection and oscillation together with the turbulent viscosity (e.g., Xiong \& Deng 2001). In the recent catalogue of $\delta$ Scuti pulsating stars, about 18 ones including VX Hya (F6) and DE Lac (F7) were reported to be beyond the red edge and thus have cool temperatures with the spectroscopic/photometric spectral types later than F5 (e.g., Chang et al. 2013).

A large number of $\delta$ Scuti stars were discovered
by several photometric surveys in the world, e.g., All Sky Automated Survey (ASAS: Pojmanski 1997; Pojmanski et al. 2005), the asteroid survey LINEAR (Palaversa et al. 2013), and northern sky variability survey (NSVS: Wo\'{z}niak et al. 2004). In particular, the Kepler (Borucki et al. 2010) space telescopes have produced high-quality data sets for thousands of stars and about 2000 $\delta$ Scuti stars were detected (e.g., Balona \& Dziembowski 2011; Uytterhoeven et al. 2011). Since the data on variable stars including $\delta$ Scuti stars are constantly varying, the mission of VSX (the international variable star index) is bringing all of new information together in a single data repository (e.g., Watson et al. 2006). About 3689 $\delta$ Scuti variable stars have been detected and were listed in VSX by 2017 July 16. Those photometric survey data are very useful to understand the photometric properties of $\delta$ Scuti variable stars. However, their spectroscopic information including spectral types and stellar atmospheric parameters extremely lacks. Among 1578 $\delta$ Scuti stars collected by Chang et al. (2013), only 15\% of them have spectroscopic spectral types.

Large Sky Area Multiobject Fibre Spectroscopic Telescope (LAMOST, also called as Guo Shou Jing telescope) is a special telescope that has a field of view of 5 deg and could simultaneously obtain the spectra of about 4000 stars with low resolution of about 1800 in a one exposure (Wang et al. 1996; Cui et al. 2012). The wavelength range of LAMOST is from 3700 to 9000\,{\AA} and is divided in two arms, i.e., a blue arm (3700-5900\,{\AA}) and a red arm (5700-9000\,{\AA}). Huge amounts of spectroscopic data have been obtained for single stars and close binaries (e.g., Zhao et al. 2012, Luo et al. 2015; Qian et al. 2017a, b). In the time interval between 2011 October 24 and 2017 June 16, about 766 $\delta$ Scuti pulsating stars were observed by LAMOST. In the recent LAMOST data releases, stellar atmospheric parameters for 525 $\delta$ Scuti stars were determined when their spectra have higher signal to noise, while spectral types of all variables were obtained. Those spectroscopic data provide valuable information on their physical properties, evolutionary states, and binarities.

Among the 766 observed $\delta$ Scuti stars, the pulsating periods of 756 samples are given in VSX. The period distribution for those observed $\delta$ Scuti variables is shown in Fig. 1. Also displayed in the figure is the period distribution of $\delta$ Scuti variable stars listed in VSX where 339 $\delta$ Scuti stars without pulsating periods are not shown.
The period distribution given by Rodr\/{i}guez \& Breger (2001) indicated that the majority of
$\delta$ Scuti stars have short periods in the range from 0.05 to 0.1\,d and the number is decreasing with increasing period.
However, as shown in Fig. 1, there are two peaks in the period distributions. The main peak is near 0.05\,d, while the the other small peak is at 0.17\,d. These properties are similar to those observed in RR Lyr-type pulsating stars (e.g., Szczygiel \& Fabrycky 2007). The latter was caused by the known two types of RR Lyr-type pulsating stars, i.e., RRab and RRc.
In the paper, we report the detection of a group of unusual and cool variable stars (hereafter UCVs) that are distinguished from normal ones. Then, the physical properties of normal $\delta$ Scuti stars (hereafter NDSTs) are investigated after UCVs are excluded. Finally, based on the distributions of those atmospheric parameters and some statistical correlations, the physical properties, the binarities and the evolutionary states of $\delta$ Scuti stars are discussed.

\section{Catalogues of spectroscopic observations for UCVs and NDSTs}

\begin{figure*}
\begin{center}
\includegraphics[width=2.2\columnwidth]{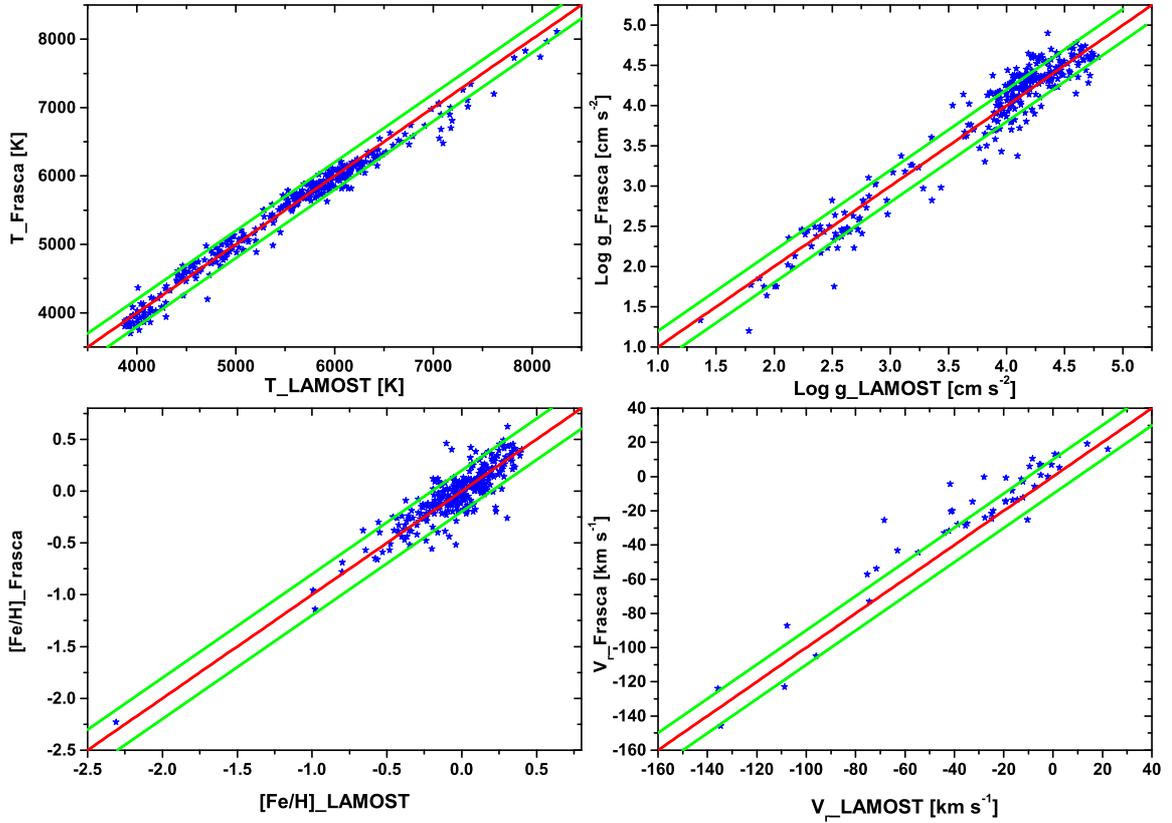}
\caption{Comparison between LASP data with literature values based mainly on high resolution spectra. The red solid lines are the one-to-one relationships. The green lines in the four panel refer to a difference of 200\,K for $T$, 0.2 for $\log g$ and [Fe/H], and 10\,km$^{-1}$ for $V_{\rm r}$, respectively.}
\end{center}
\end{figure*}

\begin{figure*}
\begin{center}
\includegraphics[width=2.2\columnwidth]{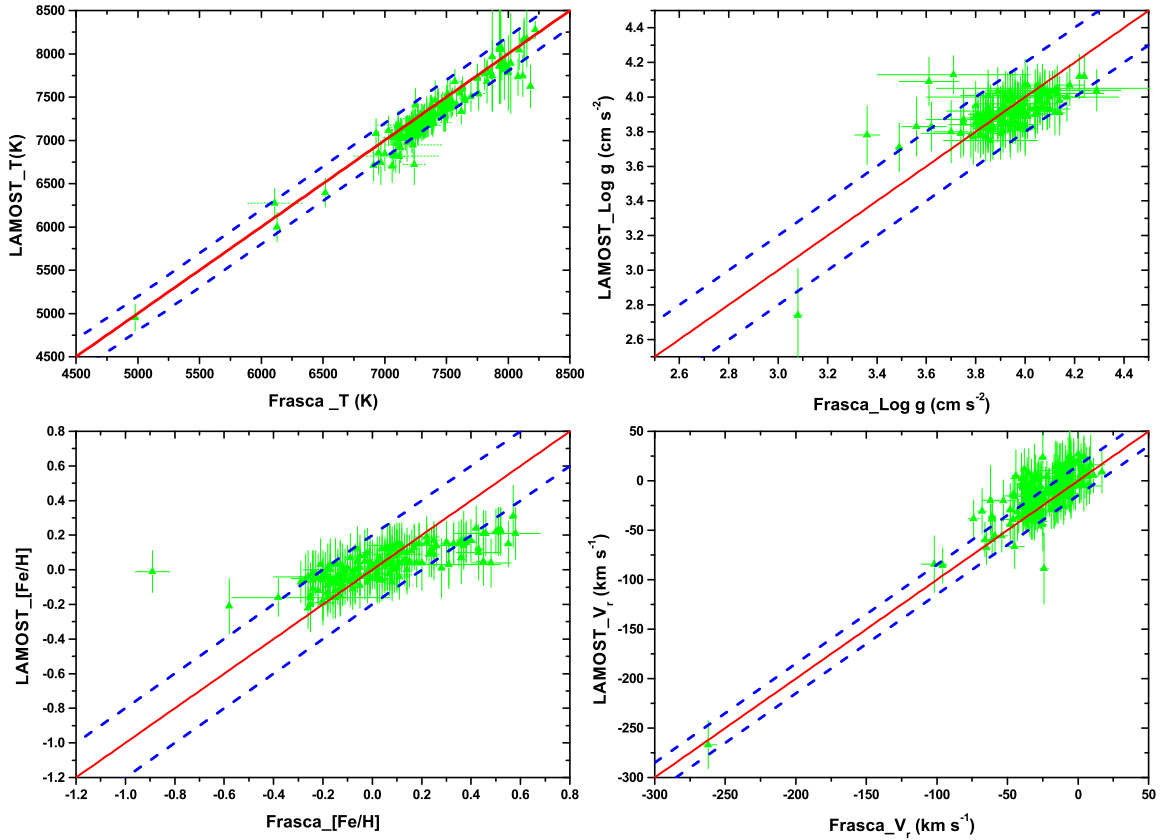}
\caption{Comparison between LASP atmospheric parameters of LAMOST-Kepler $\delta$ Scuti stars with those derived with ROTFIT from the LAMOST spectra by Frasca et al. (2016). As those in Fig. 2, the red solid lines in the four panel are the one-to-one relationships. The dashed blue lines refer to a difference of 200\,K for $T$, 0.2 for $\log g$ and [Fe/H], and 15\,km$^{-1}$ for $V_{r}$, respectively.}
\end{center}
\end{figure*}

\begin{figure}
\begin{center}
\includegraphics[width=1.15\columnwidth]{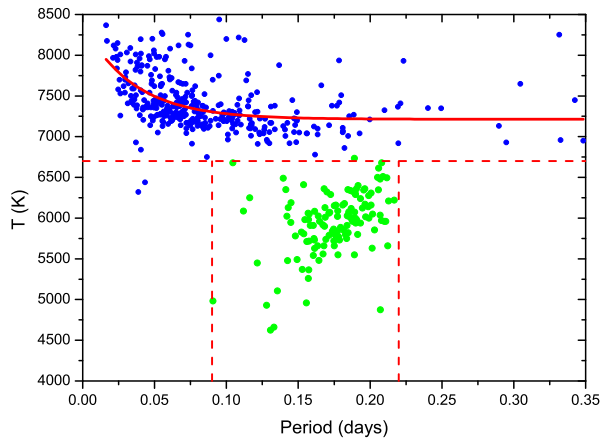}
\caption{
Correlation between the pulsating period and the effective temperature for $\delta$ Scuti variable stars with periods shorter than 0.35\,d. A group of 131 UCVs (green dots) is shown to be distinguished from the normal ones (blue dots). The dashed red lines represent their borders. The UCVs have temperatures lower than 6700\,K with pulsating periods in the range from 0.09 to 0.22\,d. The solid red line indicates that there is a good relation between the period and the effective temperature for normal $\delta$ Scuti stars with period shorter than 0.35\,d.}
\end{center}
\end{figure}

\begin{table*}
\footnotesize
\caption{Mean stellar atmospheric parameters for 32 $\delta$ Scuti stars observed 4 times or more and their standard errors.}\label{table1}
\begin{center}
\begin{tabular}{lllllllll}\hline
 Star name              &     Period (d)  &  N  &   $\overline{T}$ (K) &   Errors &   $\overline{\log g}$   &  Errors   &   $\overline{[\rm Fe/H]}$ & Errors  \\\hline
 KIC 10014548                   & 0.65445 & 11    & 7210 & 18 & 3.82 & 0.03 & 0.56 & 0.02 \\
    KIC 7697795                    & 0.057185 & 10    & 7498  & 11 & 3.86 & 0.01 & 0.08 & 0.01 \\
    KIC 7106205                    & 0.074655 & 10    & 7167 & 15 & 3.85 & 0.02 & 0.12 & 0.01 \\
    KIC 11395392                   & 0.039701 & 7     & 7261 & 45 & 4.13 & 0.02 & -0.25 & 0.02 \\
    KIC 9391395                    & 0.51308 & 7     & 7243 & 11 & 4.05 & 0.01 & -0.20 & 0.01 \\
    KIC 8149341                    & 0.037015 & 7     & 7505 & 32 & 3.84 & 0.04  & 0.16 & 0.02 \\
    KIC 3942911                    & 0.03391 & 6     & 7438 & 21 & 3.96 & 0.02 & 0.05 & 0.02 \\
    KIC 8747415                    & 0.090662 & 6     & 4989 & 30 & 3.10 & 0.04 & -0.18 & 0.03 \\
    ASAS J075941-0353.1            & 0.180982 & 6     & 7073 & 32 & 3.75 & 0.03 & 0.18 & 0.02 \\
    KID 07900367                   & 0.075489 & 5     & 7198  & 16 & 3.80 & 0.02 & 0.51 & 0.00 \\
    KIC 9324334                    & 0.097352 & 5     & 7238 & 27 & 3.96 & 0.03 & 0.13 & 0.02 \\
    KIC 10615125                   & 0.10351 & 5     & 7202 & 13 & 3.83 & 0.01 & 0.35 & 0.01 \\
    KIC 11127190                   & 0.053387 & 5     & 7448  & 16 & 3.92  & 0.01 & 0.26 & 0.01 \\
    KIC 9700679                    & 3.81679 & 5     & 5451 & 82  & 3.09 & 0.10 & 0.06 & 0.08 \\
    BW Cnc                         & 0.072 & 5     & 7421 & 40 & 3.90 & 0.02 & 0.13 & 0.01 \\
    KIC 8869892                    & 0.12989 & 4     & 7068 & 22 & 3.89 & 0.09  & -0.10 & 0.13 \\
    BD+24 95                       & 0.11992 & 4     & 7090  & 22  & 3.75 & 0.01 & 0.49 & 0.02 \\
    KIC 10002897                   & 0.064965 & 4     & 7305  & 17 & 4.06  & 0.01 & -0.10 & 0.05 \\
    GSC 01946-00035                & 0.067081 & 4     & 7265  & 40 & 4.00     & 0.01 & -0.03 & 0.03 \\
    ASAS J190751+4629.2            & 0.079561 & 4     & 7180  & 8  & 3.92 & 0.03 & 0.19 & 0.01 \\
    V0367 Cam                      & 0.121596 & 4     & 7308 & 126 & 3.86  & 0.02 & 0.19  & 0.05 \\
    ASAS J064626+2629.8            & 0.205972 & 4     & 6613 & 10  & 4.16 & 0.03 & -0.25 & 0.01 \\
    ASAS J071540+2022.2            & 0.200724 & 4     & 6308 & 5     & 4.17  & 0.01 & -0.08 & 0.02 \\
    KIC 10208345                   & 0.28969 & 4     & 7146 & 24 & 4.00 & 0.04  & -0.08 & 0.03 \\
    KIC 10065244                   & 0.083822 & 4     & 7114 & 23 & 3.97 & 0.02 & -0.03 & 0.02 \\
    KIC 9775385                    & 0.55928 & 4     & 7360  & 8  & 3.98 & 0.01 & 0.07 & 0.01 \\
    KIC 7668791                    & 0.048443 & 4     & 8197 & 12 & 3.83 & 0.01 & -0.08 & 0.01 \\
    KIC 7350486                    & 0.75188 & 4     & 7132 & 11 & 3.89 & 0.02 & 0.10 & 0.01 \\
    KIC 7109598                    & 0.053772 & 4     & 7275 & 45 & 3.87 & 0.04 & 0.36 & 0.03 \\
    KIC 5446068                    & 3.48432 & 4     & 6200  & 311 & 3.99  & 0.27 & 0.29 & 0.02 \\
    KIC 4488840                    & 0.063279 & 4     & 7275  & 26 & 3.93  & 0.01 & 0.06 & 0.02 \\
    BS Cnc                         & 0.051 & 4     & 7480  & 18 & 3.91 & 0.02 & 0.06  & 0.01 \\
\hline
\end{tabular}
\end{center}
\end{table*}

\begin{table*}
\small
\caption{Stellar atmospheric parameters for the 131 UCVs detected by LAMOST (the first 20 lines of the whole table).}\label{XXXX}
\begin{tabular}{lllllllllllll}\hline\hline
Name & R.A. (deg) & Dec. (deg) & Period (d) & Sp. & Dates & N & $T$ (K)& $E_1$ & $\log g$ & $E_2$ &[Fe/H] & $E_3$ \\\hline
     GM Com                         & 183.1038 & 27.38008 & 0.208 &     F3  &  2011-12-21--2011-12-21  & 1     & 6680  & 20    & 4.15  & 0.02  & -0.10  & 0.02  \\
     ASAS J001921+0127.1            & 4.83683 & 1.45297 & 0.156951 &     G7  &  2012-10-29--2012-10-29  & 1     & 5260  & 30    & 3.74  & 0.03  & 0.05  & 0.02  \\
     ASAS J010045+2648.4            & 15.18808 & 26.80683 & 0.190532 &     F2  &  2015-10-31--2015-10-31  & 1     & 5980  & 10    & 4.32  & -     & -0.59  & - \\
     ASAS J011159+2548.8            & 17.99496 & 25.81331 & 0.157222 &     F6  &  2014-10-25--2014-10-25  & 1     & 5980  & 290   & 3.88  & 0.41  & -0.51  & 0.27  \\
     ASAS J011208+0217.2            & 18.03254 & 2.28617 & 0.176614 &     G5  &  2012-10-31--2016-12-30  & 2     & 5761  & 5     & 4.31  & 0.01  & 0.10  & 0.01  \\
     ASAS J011959+1500.4            & 19.99579 & 15.00731 & 0.199153 &     F9  &  2016-01-19--2016-01-19  & 1     & 6000  & 20    & 4.12  & 0.02  & 0.11  & 0.01  \\
     ASAS J012334+2108.6            & 20.89288 & 21.14336 & 0.185343 &     F4  &  2012-12-03--2014-11-19  & 2     & 6020  & 55    & 4.03  & 0.07  & -0.24  & 0.05  \\
     ASAS J013458+0553.2            & 23.74217 & 5.88589 & 0.164786 &     F9  &  2015-12-11--2015-12-11  & 1     & 5960  & 320   & 4.44  & 0.45  & -0.43  & 0.29  \\
     ASAS J020653+0815.6            & 31.71892 & 8.26008 & 0.177637 &     G3  &  2012-09-29--2012-09-29  & 1     & 5550  & 100   & 3.82  & 0.14  & -0.34  & 0.09  \\
     ASAS J022414+2741.6            & 36.05646 & 27.69336 & 0.155998 &     G3  &  2013-10-06--2017-01-28  & 2     & 5713  & 5     & 4.26  & 0.01  & -0.05  & 0.00  \\
     ASAS J030054+2301.7            & 45.22333 & 23.02756 & 0.181652 &     G2  &  2015-10-30--2015-10-30  & 1     & 5850  & 90    & 4.31  & 0.12  & -0.04  & 0.08  \\
     ASAS J030701+1534.2            & 46.75504 & 15.56981 & 0.171276 &     F2  &  2012-09-30--2012-09-30  & 1     & 6060  & 20    & 4.15  & 0.02  & -0.15  & 0.01  \\
     ASAS J030858-0251.8            & 47.23971 & -2.86344 & 0.135452 &     K3  &  2017-02-06--2017-02-06  & 1     & 5106  & 42    & 4.43  & 0.06  & -0.29  & 0.04  \\
     ASAS J032314+0406.0            & 50.80875 & 4.10236 & 0.157606 &     F2  &  2014-12-02--2014-12-02  & 1     & 6049  & 100   & 4.11  & 0.14  & -0.20  & 0.09  \\
     ASAS J034144+1638.1            & 55.43454 & 16.63453 & 0.163976 &     F6  &  2016-12-19--2016-12-19  & 1     & 6027  & 7     & 4.26  & 0.01  & -0.42  & 0.01  \\
     ASAS J035613+2600.1            & 59.05558 & 26.00178 & 0.195533 &     F5  &  2016-11-10--2016-11-10  & 1     & 6345  & 2     & 4.14  & 0.00  & -0.05  & 0.00  \\
     ASAS J040216-0321.7            & 60.56533 & -3.36314 & 0.155724 &     K3  &  2012-11-01--2012-11-01  & 1     & 4960  & 60    & 4.27  & 0.08  & 0.31  & 0.05  \\
     ASAS J041458+0501.2            & 63.7415 & 5.01972 & 0.170028 &     F0  &  2015-11-29--2015-11-29  & 1     & 6320  & 10    & 4.19  & -     & -0.55  & - \\
     ASAS J052412+0020.2            & 81.05208 & 0.33606 & 0.196369 &     F5  &  2014-11-18--2014-11-18  & 1     & 6360  & 20    & 4.11  & 0.02  & -0.21  & 0.01  \\
     ASAS J053125+1103.8            & 82.85413 & 11.06303 & 0.144799 &     F5  &  2013-10-02--2014-12-10  & 2     & 5948  & 8     & 4.08  & 0.02  & -0.43  & 0.01  \\\hline
\end{tabular}
\end{table*}

\begin{table*}
\small
\caption{Catalogue of NDSTs observed by LAMOST (the first 20 observations).}\label{XXXX}
\begin{tabular}{lllllllllllll}\hline\hline
Name & R.A. (deg) & Dec. (deg) & Period (days) & Sp. & Dates & N & $T$ (K)& $E_1$ & $\log g$ & $E_2$ &[Fe/H] & $E_3$ \\\hline
     GP And                         & 13.82563 & 23.16372 & 0.078683 &     F0  &  2013-10-15--2013-10-15  & 1     & 7200  & 40    & 4.11  & 0.06  & -0.41  & 0.04  \\
     YZ Boo                         & 231.0292 & 36.86683 & 0.104092 &     A7  &  2015-02-18--2015-03-09  & 2     & 7306  & 8     & 4.05  & 0.01  & -0.29  & 0.01  \\
     BS Cnc                         & 129.7879 & 19.59239 & 0.051 &     A7  &  2015-03-08--2016-02-24  & 4     & 7471  & 6     & 3.89  & 0.01  & 0.07  & 0.01  \\
     BV Cnc                         & 130.1373 & 19.19433 & 0.21  &     A9  &  2015-10-30--2016-02-24  & 2     & 7325  & 7     & 3.95  & 0.02  & 0.23  & 0.01  \\
     BW Cnc                         & 130.2187 & 20.26653 & 0.072 &     A8  &  2015-03-08--2017-02-13  & 5     & 7421  & 4     & 3.90  & 0.00  & 0.13  & 0.00  \\
     GP Cnc                         & 129.5405 & 7.22583 &                   &     F0  &  2015-02-06--2015-02-06  & 1     & 7320  & 10    & 4.12  & 0.01  & -0.39  & 0.01  \\
     V0927 Her                      & 254.075 & 50.12664 & 0.130528 &     F0  &  2017-05-16--2017-05-16  & 1     & 7169  & 7     & 3.83  & 0.01  & 0.19  & 0.01  \\
     UX LMi                         & 161.4258 & 27.96506 & 0.15064 &     F0  &  2011-12-15--2011-12-26  & 2     & 7049  & 9     & 3.67  & 0.01  & 0.44  & 0.01  \\
     WW LMi                         & 163.6757 & 25.49072 & 0.12691 &     F0  &  2012-01-04--2012-03-11  & 2     & 7220  & 14    & 3.77  & 0.02  & 0.30  & 0.01  \\
     V0501 Per                      & 63.96912 & 51.21828 &                   &     F0  &  2012-02-08--2012-02-08  & 1     & 7140  & 140   & 4.26  & 0.19  & -0.14  & 0.12  \\
     UV Tri                         & 23.00046 & 30.36569 &                   &     F0  &  2013-10-04--2013-10-04  & 1     & 7250  & 20    & 3.89  & 0.02  & 0.14  & 0.01  \\
     V0460 And                      & 38.55942 & 42.241 & 0.074981 &     A6  &  2014-11-04--2014-11-04  & 1     & 8120  & 40    & 4.16  & 0.05  & -0.25  & 0.03  \\
     NSV 1273                       & 56.39346 & 24.46328 &                   &     A6  &  2016-09-30--2016-11-10  & 2     & 7297  & 2     & 4.08  & 0.00  & -0.10  & 0.00  \\
     V1116 Her                      & 247.5683 & 16.91833 & 0.094683 &     A7  &  2013-04-20--2013-04-20  & 1     & 7750  & 10    & 4.12  & -     & -0.41  & - \\
     NSV 12196                      & 293.8245 & 46.419 & 0.052768 &     A7  &  2012-06-04--2012-06-04  & 1     & 7430  & 10    & 3.91  & -     & 0.09  & - \\
     NSV 19942                      & 205.9122 & 21.0957 & 0.16352 &     F0  &  2015-01-23--2017-06-07  & 3     & 7099  & 22    & 3.72  & 0.03  & 0.45  & 0.02  \\
     V0478 And                      & 4.73279 & 22.66117 & 0.096 &     F0  &  2016-12-04--2016-12-04  & 1     & 7398  & 3     & 3.96  & 0.00  & -0.09  & 0.00  \\
     ASAS J012242+0854.1            & 20.67687 & 8.90192 & 0.067416 &     A7  &  2015-12-11--2015-12-11  & 1     & 7379  & 30    & 4.18  & 0.03  & -0.45  & 0.02  \\
     ASAS J024816+2213.4            & 42.06658 & 22.22214 & 0.080315 &     F0  &  2015-01-18--2015-01-18  & 1     & 7290  & 10    & 3.89  & -     & 0.01  & - \\
     ASAS J030012+0731.1            & 45.05079 & 7.51789 & 0.067219 &     A5  &  2012-11-01--2012-11-01  & 1     & 7440  & 10    & 4.14  & 0.01  & -0.44  & 0.01  \\
    \hline
\end{tabular}
\end{table*}

In the latest LAMOST data releases, about 20.8\% $\delta$ Scuti-type variables (766) listed in VSX were observed by LAMOST spectroscopic survey from 2011 October 24 to 2017 June 16. In the survey observations, every target was allocated to a fibre on the focal plane, and then be led into one of sixteen spectrographs equipped with thirty two $4\,K \times 4\,K$\,CCDs [One spectrograph is equipped with two CCDs for recording blue (3700 A-5900 A) and red (5700 A-9000 A) wavelength regions, respectively]. Five flux standard stars were assigned to each spectrograph with highest priority for data reduction. The raw CCD data were reduced by the software called LAMOST 2D pipeline (Luo et al. 2015), which task includes dark and bias subtraction, flat-field correction, spectral extraction, sky subtraction, removing of telluric absorption, wavelength and relative flux calibration, and combination of red and blue wavelength regions. The extracted 1D spectra from 2D pipeline were then processed by the LAMOST 1D pipeline for their spectral classes (star, galaxy and QSO) as well as the spectral types and radial velocities of stars, or the redshifts of galaxies and QSOs. For the spectra of star from late A to FGK type, the LASP (LAMOST Stellar Parameter Pipeline; Luo et al. 2015) is used to determine their atmospheric parameters (effective temperature $T$, surface gravity $\log g$, metallicity [Fe/H], and radial velocity $R_r$). The two child components (CFI and ULySS) in LASP will be consecutively executed in an iterative way, and the final parameters are derived by ULySS (Universite de Lyon Spectroscopic analysis Software; Koleva et al. 2009; Wu et al. 2011a) on the flux normalized spectra. The ULySS determines the atmospheric parameters by minimizing the $\chi^2$ between the observation and the model (Wu et al. 2011b). The model is cored by the TGM function which is an interpolator (Wu et al. 2011a) based on the empirical ELODIE stellar library (Prugniel \& Soubiran 2001; Prugniel et al. 2007). Meanwhile, the model also employed a Gaussian broadening function characterized by the systemic velocity $V_s$, and the dispersion $\sigma$. This dispersion was contributed by both instrumental broadening and the projected rotational velocity of stars. The star rotation is taken into account coupled with the instrumental broadening, thus the rotational velocity itself cannot be derived as an output parameter.

Several independent works on comparing parameters between the earlier LAMOST data release and other reliable spectral data bases, such as high resolution spectral results, SDSS, APOGEE, and PASTEL, were carried out and described by Luo et al (2015). The typical standard deviations from these comparisons are 95\,K for $T$, 0.25 dex for $\log g$, 0.1 dex for [Fe/H], and 7\,km/s for $R_r$. For the latest LAMOST data (DR4 and the first three quarters of DR5) used in this work, the same comparisons are required. To this aim we compared the LASP data with those mainly derived from high-resolution optical spectra from the literature by Frasca et al. (2016). The comparisons are plotted in the four panels of Fig. 2 where the stellar atmospheric parameters of 352 stars are shown. We note that there are very good agreements for those four parameters. The dashed blue lines in those panels refer to a difference of 200\,K for $T$, 0.2 for $\log g$ and [Fe/H], and 15\,km/s for $V_{r}$, respectively. Most of the targets are within the differences. The standard deviations are derived as 135\,K for $T$, 0.21 dex for $\log g$, 0.14 dex for [Fe/H], and 11\,km/s for $V_{r}$.

About 2000 $\delta$ Scuti variables were found by the Kepler space telescope (e.g., Balona \& Dziembowski 2011; Uytterhoeven et al. 2011). Some of them were also observed by LAMOST from 2011 October 24 to 2017 June 16. They were included in the LAMOST-Kepler project (e.g., Frasca et al. 2016; Gray et al. 2016; Ren et al. 2016). Based on the LAMOST spectra, the atmospheric parameters for tens of thousands of stars in the Kepler field were determined by Frasca et al. (2016) with the code ROTFIT that was developed by Frasca et al. (2003, 2006). Stellar atmospheric parameters of 188 $\delta$ Scuti stars were obtained. The comparisons between the present atmospheric parameters of the 188 LAMOST-Kepler $\delta$ Scuti stars with those derived with ROTFIT are show in Fig. 3 where the red solid lines in the four panel represent the one-to-one relationships. The dashed blue lines refer to a difference of 200\,K for $T$, 0.2 for Log g and [Fe/H], and 15\,km/s for $V_{r}$, respectively. It is shown that there is a good agreement for the effect temperature. As for Log g, apart from a few targets, most $\delta$ Scuti stars are within the difference. The [Fe/H] values are only in good agreement with the ROTFIT data when $[Fe/H] < 0.3$. As for the radial velocity, most $\delta$ Scuti stars are also within the difference of 15\,km/s although the ROTFIT data tend to have lower values for some targets.

The temperatures of all targets are below 8500\,K, their atmospheric parameters could be determined well. Among the 766 $\delta$ Scuti stars, stellar atmospheric parameters of 525 $\delta$ Scuti stars were determined by using good and reliable spectra. Their atmospheric parameters were automatically determined by ULySS (e.g., Koleva et al. 2009; Wu et al. 2011a, b, 2014; Luo et al. 2015). There are 32 $\delta$ Scuti variable stars observed four times or more on different dates. To check the reliability of stellar atmospheric parameters, we determined the mean values of their atmospheric parameters and derived the corresponding standard errors. The results are shown in Table 1, where their names and pulsating periods are listed in the first and the second columns. The observational times are shown in third column, while the average atmospheric parameters and their standard errors are displayed in the rest columns. As shown in Table 1, apart from two targets, the standard errors of the effective temperature for the rest targets are lower than 100\,K. As for the gravitational acceleration $\log g$, apart from one target, the standard errors of the others are lower than 0.1\,dex, while the standard errors of the metallicity for most $\delta$ Scuti variable star targets are lower than 0.08\,dex. These results are consistent with the standard deviations determined by comparing LAMOST data with those derived from high-resolution spectra. This indicates that the present data for those $\delta$ Scuti stars are reliable.

The correlation between the pulsating period $P$ and the effective temperature $T$ for those observed $\delta$ Scuti stars is displayed in Fig. 4. The pulsating periods in the figure are from VSX\footnote{http://www.aavso.org/vsx/} (e.g., Watson, 2006). As shown in Fig. 4, there is a group of 131 UCVs (green dots) that are clearly separated from the other 394 variable stars (blue dots) indicating that they are a new group of unusual variable stars. The dashed red lines in the figure represent the borders of UCVs. Their temperatures are lower than 6700\,K and the periods are in the range from 0.09 to 0.22\,d. Four targets, KIC\,9700679 (P=3.81679\,d), KIC\,5446068 (P=3.48432\,d), BOKS-24178 (P=0.7\,d), and $BD\,+19^{\circ}572$ (without period), with temperatures below 6700\,K are not shown in the figure because their periods are not in the range. The atmospheric parameters of the 131 UCVs including the effective temperature T, the surface gravity Log g, and the metallicity [Fe/H] are listed in Table 2 in the order of increasing VSX number.
When some of them were observed twice or more times on different dates, the atmospheric parameters are averaged and the corresponding standard errors are given.
Those listed in Table 2 are the first 20 sets of observations. The whole table is available in the internet electronic version\footnote{http://search.vbscn.com/DSCT.Table2.txt} and it will be improved by adding new data obtained by LAMOST in the future. Table 2 includes star names, their right ascensions ($RA$) and declinations ($DEC$), and pulsating periods. These parameters are from VSX catalogue. When the targets are observed two times or more, the effective temperature $T$, the gravitational acceleration $\log g$ and the metallicity [Fe/H] are averaged and the weight of each value is the inverse square of the original error.
The mean spectral types are listed in columns 5 and those shown in columns 6 and 7 are observational dates and times (N). Only the first date and the last date are given in the table as the targets were observed more than two times. The stellar atmospheric parameters including $T$, $\log g$, and [Fe/H], are listed in columns 8, 10, and 12, while their corresponding errors are displayed in columns 9, 11, and 13, respectively. The stellar atmospheric parameters of the other 394 NDSTs are shown in Table 3 in the order of increasing VSX number. The arrangement of the table is the same as those in Table 2 where only the first 20 observations are shown in Table 3 (the whole table is available via the internet electronic version\footnote{http://search.vbscn.com/DSCT.Table3.txt}).

For some $\delta$ Scuti stars, their spectra signals to noise are not high enough to determine the stellar atmospheric parameters and thus only spectral types were given. The spectral types of those $\delta$ Scuti variable stars are catalogued in an electronic table where 588 spectral types are listed. This table is available at the website\footnote{http://search.vbscn.com/DSCT.LAMOSTdata.txt} via the internet. For about 241 $\delta$ Scuti stars, only spectral type was obtained by LAMOST. Both spectral types and stellar atmospheric parameters were determined for 525 $\delta$ Scuti stars. In order to provide more information to the readers, the original LAMOST stellar parameter for UCVs and NDSTs are also included in the table. In total, 182 sets of spectroscopic observations for 131 UCVs and 661 sets of data for 394 NDSTs are also catalogued in this table that contains 16 columns. As those shown in Tables 2 and 3, the first three columns include the star names, their right ascensions $RA$ and declinations $DEC$. The types of light variation and pulsating periods are listed in columns 4 and 5. These parameters are from VSX catalogue. The types of light variation are defined in the GCVS variability classification scheme\footnote{http://www.sai.msu.su//gcvs/gcvs/iii/vartype.txt} (e.g., Samus et al. 2017).  Column 6 lists the distances (in arcsec) between the two positions determined by the coordinates given in VSX and by LAMOST. As pointed by Qian et al. (2017a), the distances were used to identify those $\delta$ Scuti variable stars from the LAMOST samples based on the criterion Dist$<2$\,arcsec.  For most of the targets, the values of Dist are smaller than 0.1\,arcsecs. However, the Dist values of several $\delta$ Scuti are larger than 1\,arcsec. We have checked their stellar charts by using the coordinates given in VSX and by LAMOST. It is shown that they are the same stars and their results are not for different objects. The differences may be caused by the fact that the coordinates for a few targets are not given in high precision. The observing dates are shown in column 7, while the determined spectral types are listed in columns 8. The stellar atmospheric parameters including $T$, $\log g$, [Fe/H], and $V_{\rm r}$), are listed in columns 9, 11, 13, and 15. Their corresponding errors $E_1$, $E_2$, $E_3$, and $E_4$ are also displayed in the table, respectively.

\section{Distributions of stellar atmospheric parameters and Binarities for $\delta$ Scuti variables}

\begin{figure*}
\begin{center}
\includegraphics[width=1.15\columnwidth]{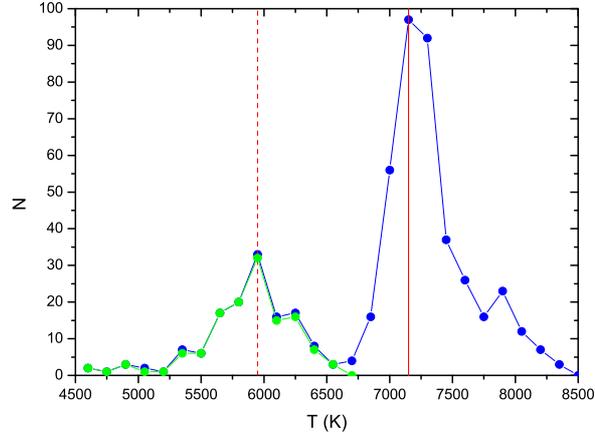}
\caption{Distribution of the effective temperature for whole $\delta$ Scuti variable stars whose stellar atmospheric parameters were determined by LAMOST (blue dots). The solid red line refers to the main peak near 7150\,K, while the dashed red line to the small peak near 5950\,K. The temperature distribution for the 131 UCVs (green dots) is also potted. It is shown that the small peak is constructed by the existence of the group of cool variable stars.}
\end{center}
\end{figure*}

\begin{figure}
\begin{center}
\includegraphics[width=1.15\columnwidth]{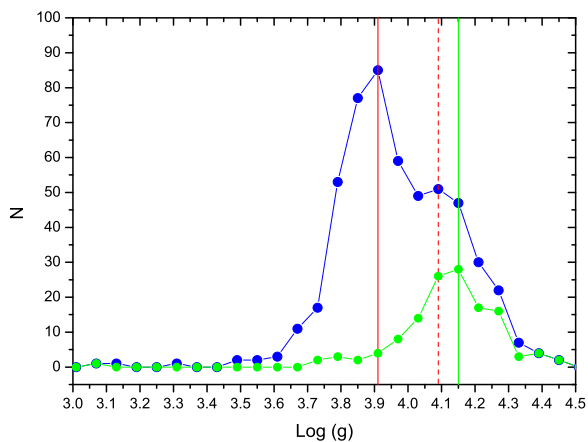}
\caption{Distribution of the gravitational acceleration $\log g$. The solid and the dashed lines represent to the two peaks near 3.91 and 4.09, respectively. The green line refers to the peak for UCVs. Symbols are the same as those in Fig. 5.}
\end{center}
\end{figure}

\begin{figure}
\begin{center}
\includegraphics[width=1.15\columnwidth]{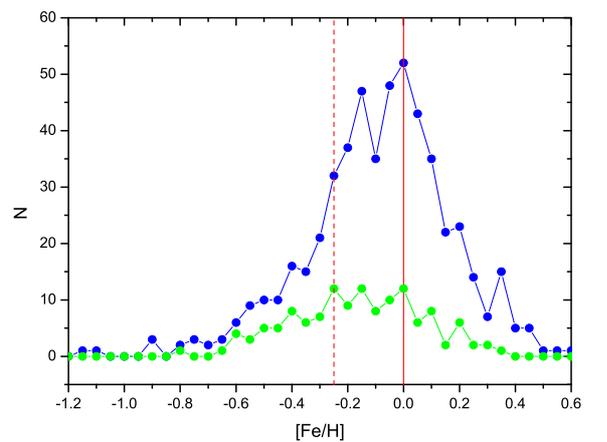}
\caption{Distribution of the metallicity [Fe/H] for $\delta$ Scuti variables observed by LAMOST. It is shown that most of the $\delta$ Scuti stars have metallicity close to the Sun, i.e., $[\rm Fe/H] \sim 0$ (the solid line), while the metallicity of most UCVs are mainly from about -0.25 (the dashed line) to 0.0. Symbols are the same as those in Figs 5 and 6.}
\end{center}
\end{figure}

\begin{table*}
\footnotesize
\begin{center}
\caption{10 $\delta$ Scuti variable stars with the highest metallicities.}\label{XXXX}
\begin{tabular}{lllllllll}\hline\hline
Name &  $\alpha(2000)$ & $\delta(2000)$ & Period (d)  & Sp.  & $T$ (K)& $\log g$ &[Fe/H] & $V_{\rm r}$ (km s$^{-1}$) \\\hline
    V2703 Cyg                      & 302.0067 & 30.87896 & 0.1166   &             F0  & 7100 & 3.74  & 0.60 & -3 \\
    KIC 10014548                   & 293.0593 & 46.90581 & 0.65445  &             F0  & 7206 & 3.83 & 0.56 & -64 \\
    KID 07900367                   & 294.9373 & 43.69592 & 0.075489 &             F0  & 7198    & 3.8   & 0.51  & -24 \\
    KIC 8103917                    & 294.0302 & 43.91889 & 0.056348 &             F0  & 7310    & 3.82  & 0.49  & -27 \\
    KIC 9874181                    & 282.8607 & 46.78    & 0.049222 &             F0  & 7440    & 3.81  & 0.49  & -29 \\
    BD+24 95                       & 10.26    & 25.26081 & 0.11992  &             F0  & 7092    & 3.75  & 0.49  & -20\\
    KIC 4936524                    & 295.2013 & 40.04136 & 0.035665 &             F0  & 7391    & 3.83  & 0.46 & -7 \\
    NSV 19942                      & 205.9122 & 21.0957  & 0.16352  &             F0  & 7099    & 3.72  & 0.45 & 6 \\
    KIC 5722346                    & 296.3865 & 40.94497 & 0.0883   &             F0  & 7130    & 3.90  & 0.45  & -15 \\
    UX LMi                         & 161.4258 & 27.96506 & 0.15064  &             F0  & 7050    & 3.67  & 0.44  & -8 \\

\hline\hline
\end{tabular}
\end{center}
\end{table*}

\begin{table*}
\footnotesize
\begin{center}
\caption{50 $\delta$ Scuti pilsating stars with the difference of radial velocity larger than 15\,km s$^{-1}$.}\label{XXXX}
\begin{tabular}{llllllllll}\hline\hline
Name &  Period (d) & Dates  & Times & $T$ (K) & $\log g$ & [Fe/H] & $V_{\rm Low}$ &  $V_{\rm High}$ &  $\Delta{V}$ \\\hline
     KIC 5722346                   & 0.0883 &   2013-10-05--2015-10-08  & 3     & 7130  & 3.90  & 0.45  & -53   & 11    & 64 \\
     ASAS J081813-0049.8           & 0.139603 &   2013-11-25--2015-02-05  & 2     & 7129  & 3.80  & 0.21  & -47   & 11    & 58 \\
     KIC 10014548                  & 0.65445 &   2012-06-04--2017-06-14  & 11    & 7206  & 3.83  & 0.56  & -102  & -50   & 52 \\
     ASAS J085536+0215.5           & 0.207902 &   2012-02-10--2016-04-03  & 3     & 5970  & 4.30  & -0.36  & -25   & 18    & 43 \\
     ASAS J022414+2741.6           & 0.155998 &   2013-10-06--2017-01-28  & 2     & 5713  & 4.26  & -0.05  & -85   & -44.45 & 40.55 \\
     ASAS J154613-0026.1           & 0.068991 &   2012-05-16--2016-05-18  & 3     & 7458  & 4.05  & -0.37  & -10   & 27    & 37 \\
     NSVS 11000901                 & 0.39787221 &   2017-04-22--2017-06-11  & 2     & 6899  & 4.10  & -0.20  & -49.3 & -13.65 & 35.65 \\
     KIC 4252757                   & 0.046555 &   2012-06-15--2014-06-02  & 2     & 7541  & 3.95  & -0.09  & -68   & -34   & 34 \\
     KIC 7350486                   & 0.75188 &   2013-09-26--2017-06-15  & 4     & 7127  & 3.89  & 0.10  & -59   & -25   & 34 \\
     KIC 3634384                   & 0.085852 &   2012-06-15--2014-06-02  & 3     & 7360  & 3.87  & 0.10  & -5    & 27    & 32 \\
     BS Cnc                        & 0.051 &   2015-03-08--2016-02-24  & 4     & 7471  & 3.89  & 0.07  & 5     & 36    & 31 \\
     KIC 5476864                   & 0.59988 &   2013-10-05--2014-05-22  & 2     & 7068  & 4.08  & 0.09  & -38   & -9    & 29 \\
     KID 4252716                   & 0.076184 &   2012-06-15--2017-06-03  & 2     & 7505  & 3.86  & -0.06  & -40   & -11.89 & 28.11 \\
     KIC 3942911                   & 0.03391 &   2012-06-15--2015-10-12  & 6     & 7438  & 3.96  & 0.05  & -39   & -11   & 28 \\
     ASAS J125114+0917.7           & 0.182872 &   2014-01-13--2017-05-17  & 2     & 5795  & 4.18  & 0.03  & -14   & 13.9  & 27.9 \\
     NSVS 3820963                  & 0.16784703 &   2014-11-04--2015-12-28  & 2     & 5660  & 4.26  & -0.13  & -98   & -71   & 27 \\
     BD+24 95                      & 0.11992 &   2012-09-29--2013-12-19  & 4     & 7092  & 3.75  & 0.49  & -33   & -7    & 26 \\
     KIC 9391395                   & 0.51308 &   2014-05-02--2017-06-15  & 7     & 7236  & 4.05  & -0.19  & -30   & -4    & 26 \\
     KIC 11193046                  & 0.78431 &   2013-05-22--2017-05-16  & 3     & 7993  & 3.92  & -0.12  & -25   & 0.6   & 25.6 \\
     KIC 8429756                   & 0.036041 &   2013-05-19--2015-10-11  & 2     & 7457  & 3.85  & -0.06  & -31   & -6    & 25 \\
     LINEAR 9272851                & 0.080104 &   2016-01-13--2016-02-06  & 2     & 7192  & 4.26  & -0.88  & 249   & 273   & 24 \\
     ASAS J075428+1407.8           & 0.189175 &   2013-02-15--2013-12-14  & 3     & 6073  & 4.29  & -0.41  & 29    & 53    & 24 \\
     ASAS J113810+1905.1           & 0.153786 &   2015-12-21--2017-05-15  & 2     & 5806  & 3.97  & -0.21  & -35.13 & -12   & 23.13 \\
     V0367 Cam                     & 0.121596 &   2012-01-13--2016-01-19  & 4     & 7307  & 3.84  & 0.26  & -16   & 7     & 23 \\
     KIC 5768203                   & 0.12807 &   2013-10-07--2017-06-07  & 2     & 6962  & 4.28  & -0.59  & -33   & -10.27 & 22.73 \\
     KIC 9072011                   & 0.16351 &   2014-05-29--2017-05-14  & 2     & 7026  & 3.69  & 0.23  & -34   & -13.08 & 20.92 \\
     NSV 19942                     & 0.16352 &   2015-01-23--2017-06-07  & 3     & 7099  & 3.72  & 0.45  & -2    & 18.27 & 20.27 \\
     KIC 3119604                   & 0.024301 &   2012-06-15--2017-06-13  & 2     & 7940  & 4.20  & -0.26  & -50   & -30.42 & 19.58 \\
     KIC 9450940                   & 0.033337 &   2014-05-29--2015-10-06  & 3     & 7922  & 4.08  & -0.20  & -61   & -42   & 19 \\
     KIC 7106205                   & 0.074655 &   2013-05-19--2017-06-15  & 10    & 7173  & 3.86  & 0.13  & -29   & -10   & 19 \\
     ASAS J064626+2629.8           & 0.205972 &   2012-03-11--2016-01-18  & 4     & 6613  & 4.19  & -0.25  & 0     & 19    & 19 \\
     ASAS J083952+0052.4           & 0.133043 &   2013-03-24--2013-03-24  & 2     & 4660  & 4.37  & 0.03  & 51    & 70    & 19 \\
     KID 6593488                   & 0.10101 &   2015-09-25--2016-05-18  & 3     & 7026  & 3.96  & -0.09  & -42   & -24   & 18 \\
     BD+38 2361                    & 0.049839 &   2014-01-12--2016-01-28  & 2     & 7411  & 3.97  & -0.08  & -34   & -16   & 18 \\
     KIC 8460993                   & 0.05535 &   2013-09-25--2017-06-14  & 3     & 7193  & 3.90  & 0.25  & -68   & -50   & 18 \\
     KIC 9700679                   & 3.81679 &   2013-10-04--2017-06-15  & 5     & 5491  & 3.17  & 0.10  & -23   & -5.06 & 17.94 \\
     KIC 9473000                   & 0.11496 &   2012-06-04--2017-05-16  & 2     & 7376  & 3.99  & 0.04  & -31.03 & -14   & 17.03 \\
     GSC 01946-00035               & 0.06708077 &   2013-12-01--2014-12-02  & 4     & 7267  & 4.00  & -0.03  & 59    & 76    & 17 \\
     ASAS J190751+4629.2           & 0.079561 &   2013-10-04--2015-05-30  & 4     & 7174  & 3.91  & 0.19  & -71   & -54   & 17 \\
     NSVS 9420734                  & 0.19695349 &   2014-09-17--2016-02-10  & 3     & 6015  & 3.97  & 0.04  & -20   & -3    & 17 \\
     ASAS J163007+0125.2           & 0.160533 &   2013-03-24--2013-03-24  & 2     & 5542  & 4.26  & -0.26  & -33   & -16   & 17 \\
     KIC 5272673                   & 0.059492 &   2012-06-17--2017-06-13  & 3     & 7156  & 3.95  & 0.38  & -45   & -28.26 & 16.74 \\
     KID 2304168                   & 0.123344 &   2012-06-15--2017-06-13  & 2     & 7222  & 3.86  & 0.09  & -32   & -15.42 & 16.58 \\
     HAT 199-00623                 & 0.29481 &   2012-06-17--2015-10-18  & 2     & 6930  & 3.49  & 0.42  & -75   & -59   & 16 \\
     KIC 9052363                   & 0.023829 &   2013-09-25--2015-10-03  & 3     & 7840  & 4.08  & -0.15  & -32   & -16   & 16 \\
     KIC 9156808                   & 0.046995 &   2014-09-13--2017-06-14  & 3     & 7139  & 4.11  & -0.03  & -26   & -10.56 & 15.44 \\
     ASAS J093944+0010.3           & 0.196522 &   2012-02-16--2015-01-03  & 2     & 6170  & 4.15  & -0.22  & 29    & 44    & 15 \\
     NSVS 9431318                  & 0.17746757 &   2013-02-09--2014-11-01  & 3     & 6089  & 4.08  & -0.56  & 49    & 64    & 15 \\
     1SWASP J032748.43+343810.3    & 0.11771 &   2013-09-14--2014-11-01  & 2     & 7166  & 3.95  & -0.25  & -30   & -15   & 15 \\
     KIC 7977996                   & 0.090662 &   2013-10-05--2015-10-08  & 2     & 7318  & 4.14  & -0.11  & 2     & 17    & 15 \\
\hline\hline
\end{tabular}
\end{center}
\end{table*}

\begin{table*}
\footnotesize
\begin{center}
\caption{38 $\delta$ Scuti variable stars with radial velocities higher than 70\,km s$^{-1}$.}\label{XXXX}
\begin{tabular}{lllllll}\hline\hline
Name &  Period (d)  & Sp.  & $T$ (K)& $\log g$ &[Fe/H] & $V_{\rm r}$ (km s$^{-1}$) \\\hline
    LINEAR 9272851                 & 0.080104 &        A7V  & 7060  & 4.23  & -0.94  & 273 \\
    KID 8004558                    & 0.042706 &        A5V  & 7450  & 4.29  & -0.89  & -262 \\
    LINEAR 9272851                 & 0.080104 &        A7V  & 7220  & 4.27  & -0.87  & 249 \\
    LINEAR 23008936                & 0.060405 &        A5V  & 7320  & 4.24  & -0.74  & 200 \\
    LINEAR 565638                  & 0.086365 &       A7IV  & 6750  & 4.23  & -0.74  & -177 \\
    LINEAR 19119847                & 0.080278 &        A7V  & 7330  & 4.16  & -0.27  & -122 \\
    KIC 10014548                   & 0.65445 &         F0  & 7230  & 3.74  & 0.58  & -102 \\
    LINEAR 3753972                 & 0.064807 &       A6IV  & 7511  & 4.29  & -0.72  & 101.18 \\
    Renson 31680                   & 0.46666 &        A7V  & 7570  & 4.12  & -0.36  & -99 \\
    NSVS 3820963                   & 0.167847 &         G4  & 5740  & 4.29  & -0.12  & -98 \\
    KIC 10014548                   & 0.65445 &         F0  & 7230  & 3.84  & 0.57  & -96 \\
    Renson 31680                   & 0.46666 &        A7V  & 7540  & 4.10  & -0.37  & -95 \\
    Renson 31680                   & 0.46666 &        A7V  & 7480  & 4.12  & -0.36  & -94 \\
    ASAS J053125+1103.8            & 0.144799 &         F2  & 5970  & 4.23  & -0.33  & 88 \\
    CR Ari                         & 0.18094 &        A2V  & 7110  & 3.92  & -0.79  & -87 \\
    ASAS J022414+2741.6            & 0.155998 &         G3  & 5690  & 4.23  & -0.06  & -85 \\
    ASAS J152257+1054.4            & 0.054955 &        A5V  & 8280  & 4.07  & -0.31  & -83 \\
    ASAS J053125+1103.8            & 0.144799 &         F8  & 5860  & 4.08  & -0.43  & 82 \\
    ASAS J170044-0241.3            & 0.160861 &         G2  & 5750  & 3.99  & -0.26  & -80 \\
    ASAS J135248+1150.2            & 0.186274 &         F5  & 6340  & 4.23  & -0.47  & -79 \\
    ASAS J151411+2043.1            & 0.212255 &         G4  & 5660  & 4.09  & 0.04  & -79 \\
    V0460 And                      & 0.074981 &       A6IV  & 8120  & 4.16  & -0.25  & -77 \\
    ASAS J084602+1301.4            & 0.121261 &         F0  & 7340  & 3.93  & 0.12  & 77 \\
    KIC 10014548                   & 0.65445 &         F0  & 7230  & 3.82  & 0.59  & -77 \\
    GSC 01946-00035                & 0.067081 &         F0  & 7210  & 3.98  & 0.01  & 76 \\
    HAT 199-00623                  & 0.29481 &         F0  & 7030  & 3.59  & 0.43  & -75 \\
    CSS\_J085942.7+190007           & 0.180998 &         F2  & 6160  & 4.30  & -0.49  & -75 \\
    KIC 5038228                    & 1.10619 &         F0  & 7250  & 3.62  & 0.45  & -74 \\
    GSC 01732-00480                & 0.0606 &        A7V  & 7701  & 4.16  & -0.38  & -72.77 \\
    V1116 Her                      & 0.094683 &        A7V  & 7750  & 4.12  & -0.41  & -72 \\
    ASAS J084602+1301.4            & 0.121261 &         F0  & 7570  & 3.92  & 0.08  & 72 \\
    VSX J175138.3+373909           & 0.04038 &        A7V  & 7570  & 3.86  & 0.11  & -72 \\
    ASAS J155047+0505.4            & 0.15218 &         G3  & 5730  & 4.19  & -0.18  & -71 \\
    ASAS J190751+4629.2            & 0.079561 &         F0  & 7190  & 3.93  & 0.17  & -71 \\
    NSVS 3820963                   & 0.167847 &         G3  & 5660  & 4.26  & -0.13  & -71 \\
    Galati V3                      & 0.098 &         F0  & 7190  & 3.96  & 0.06  & -71 \\
    ASAS J083952+0052.4            & 0.133043 &         K5  & 4550  & 4.36  & 0.04  & 70 \\
    ASAS J180757+0901.8            & 0.10531 &       A6IV  & 7270  & 3.96  & 0.02  & -70 \\
\hline\hline
\end{tabular}
\end{center}
\end{table*}

\begin{figure}
\begin{center}
\includegraphics[width=1.15\columnwidth]{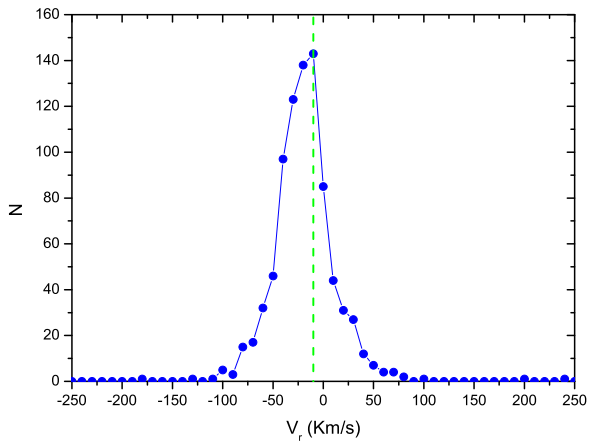}
\caption{Distribution of the radial velocity $V_{\rm r}$ for $\delta$ Scuti pulsating stars observed by LAMOST. There is a peak near $V_{\rm r}=-10$\,km s$^{-1}$.}
\end{center}
\end{figure}

\begin{figure}
\begin{center}
\includegraphics[width=1.15\columnwidth]{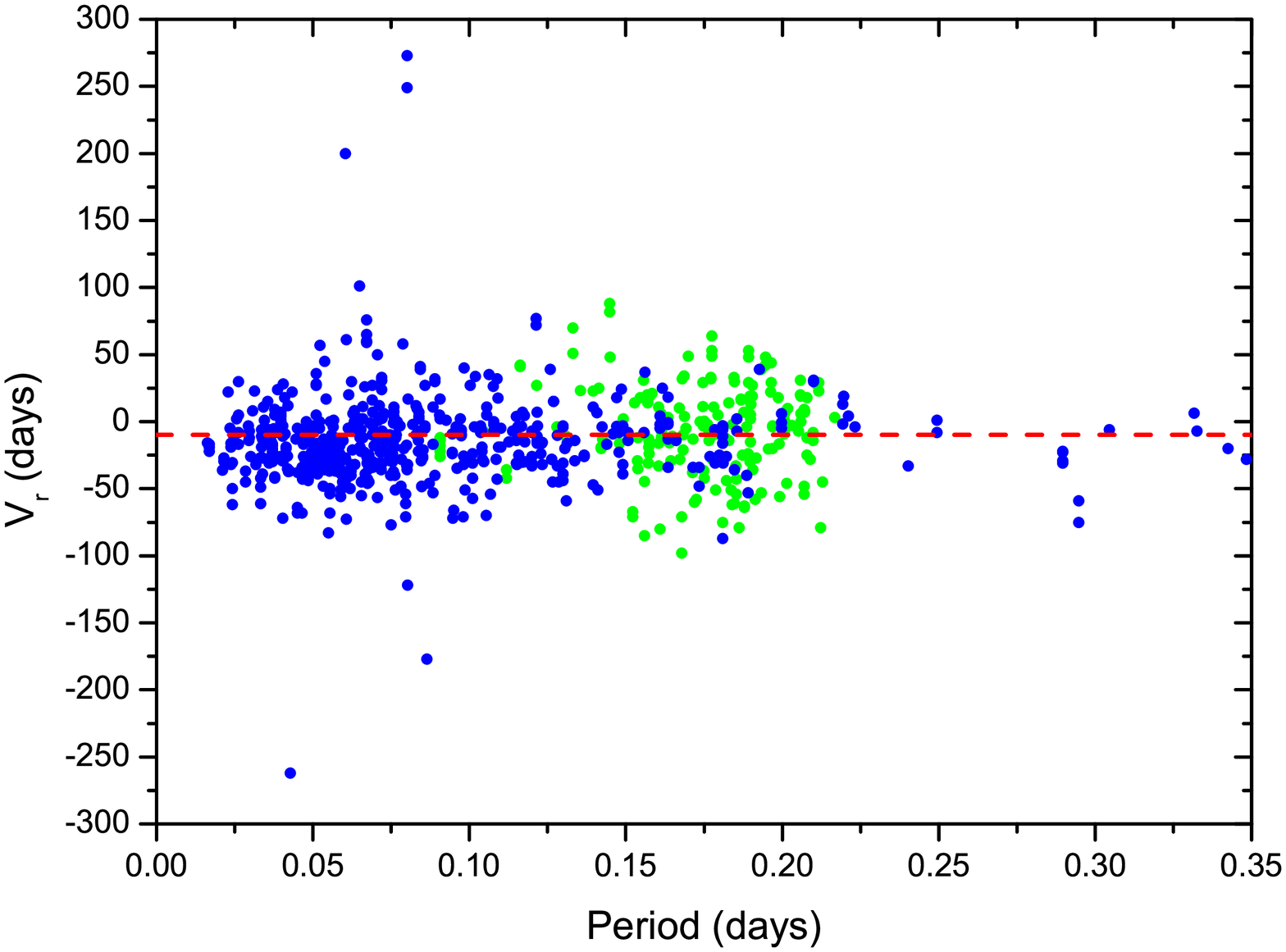}
\caption{The correlation between the radial velocity and the pulsating period. Blue dots refer to NDSTs, while green dots to UCVs (see text for details). The red dashed line represents the peak of the distribution peak of the radial velocity.}
\end{center}
\end{figure}

As aforementioned, the stellar atmospheric parameters of 525 $\delta$ Scuti pulsating stars were determined. In this section, we investigate the properties of the $\delta$ Scuti stars by using those spectroscopic data listed in Tables 2 and 3. During the analyses, when the $\delta$ Scuti variables were observed two times or more by LAMOST, the weighted mean values of all individuals are used and the weight is the inverse square of the original error from each observation. As for the radial velocity $V_{\rm r}$, we do not average them and use the individual values. The distribution of the effective temperature for 525 $\delta$ Scuti stars observed by LAMOST is shown in Fig. 5 as blue dots. As displayed in the figure, there are two peaks in the distribution. The main peak is at 7150\,K, while the other small peak is near 5950\,K. Also shown in the figure as green dots is the temperature distribution for the 131 UCVs. It suggests that the small peak is caused by the existence of the UCVs indicating that they are solar-type cool stars.

The distribution of the gravitational acceleration $\log g$ is plotted in Fig. 6. The main peak of the distribution is near 3.91. Apart from the main peak, there is a small peak at 4.09. These properties reveal that $\delta$ Scuti variable stars populate on the instability strip slightly above the zero-age main sequence (ZAMS), the main sequence, and the early stages of H-shell burning. The green dots in the figure represent the distribution of the gravitational acceleration $\log g$ for UCVs where the peak is at 4.15 indicating that their gravitational accelerations are higher than the others. The metallicity ([Fe/H]) distribution is shown in Fig. 7. The peak of the distribution is near $[\rm Fe/H] \sim 0$ indicating that most of the $\delta$ Scuti variables have metallicity close to that of the Sun. 10 $\delta$ Scuti stars have the highest metallicities are shown in Table 4. As those in Figs 5 and 6, the green dots in Fig. 7 refer to the distribution for UCVs.
It is shown that most UCVs have Fe/H from about -0.25 to 0.0 revealing that they are slightly metal poor than the normal $\delta$ Scuti stars.

The distribution of the radial velocity ($V_{\rm r}$) for those $\delta$ Scuti variable stars is displayed in Fig. 8. 843 RVs for 525 $\delta$ Scuti variable stars were determined by LAMOST from 2011 October 24 to 2017 June 16 and they are used for constructing the figure. A peak is near $V_{\rm r}=-10$\,km s$^{-1}$ indicating that the $V_0$ of most variable stars are close to this value. The relation between the pulsating period and the radial velocity is displayed in Fig. 9 where the dashed line refers to the distribution peak of the radial velocity. Those radial velocities of $\delta$ Scuti stars are useful to investigate their binarities. It is known that about 60\%$-$80\% of field stars in the solar neighbourhood are members of binary or multiple systems (e.g., Duquennoy Mayor 1991). A very interesting question is how much percent for $\delta$ Scuti stars are in binaries or multiples? Gravitational effects of close binary companions may be important to influence the non-radial pulsations through tidal interactions (e.g., Szatmary 1990). In the catalogue published by Rodr\'{i}guez et al. (2000), 86 $\delta$ Scuti stars were pointed out as members of binary or multiple systems. Among 1578 $\delta$ Scuti stars in the most recent catalogue published by Chang et al. (2013), 141 cases belong to binaries or multiples. Some of the binary $\delta$ Scuti systems were detected by analysing the light-travel time effect (e.g., Li et al. 2010, 2013; Qian et al. 2015).

Among binary $\delta$ Scuti systems, there is an interesting group of eclipsing binaries that are commonly referred to as ¡®oEA stars¡¯ (i.e., oscillating
eclipsing systems of Algol type). They are a new class of stellar systems where (B)A-F type primaries are the mass-accreting
$\delta$ Scuti stars (e.g., Mkrtichian et al. 2002; Liakos et al. 2012). Physical properties of those binary $\delta$ Scuti stars were investigated by some investigated (e.g., Mkrtichian et al. 2003; Soydugan et al. 2006; Liakos \& Niarchos 2017; Kahraman et al. 2017). Because of the pulsation, the radial-velocity changes should be associated with the light variations of $\delta$ Scuti stars. However, the information on the radial velocity extremely lacks.

Among the 525 $\delta$ Scuti pulsating stars derived spectroscopic parameters, 178 were observed two times or more by LAMOST. The difference between the lowest and the highest radial velocities are determined. 50 $\delta$ Scuti stars with $V_{\rm r}$ difference larger than 15\,km s$^{-1}$ are listed in Table 5. The peak-to-peak radial-velocity amplitude for $\delta$ Scuti stars is usually in the range from 5 to 10\,km s$^{-1}$ (e.g., Breger et al. 1976; Yang \& Walker 1986). In previous section, the LAMOST data have been compared with those determined by using high-resolution optical spectra collected from the literature by Frasca et al. (2016). It is shown that the standard deviation for radial velocity $V_{r}$ is $11$\,km s$^{-1}$. There those $\delta$ Scuti stars with $\Delta{V_{\rm r}}>15$\,km s$^{-1}$ may be the candidates of binary or multiple systems. Moreover, as shown in Fig. 9, the radial velocities for most $\delta$ Scuti stars are around $V_{\rm r}=-10$\,km s$^{-1}$. However, the radial velocities of some $\delta$ Scuti variables are very high. 38 $\delta$ Scuti stars with $V_{\rm r}>70$\,km s$^{-1}$ are shown in Table 6. They are also candidates of binary $\delta$ Scuti systems.

\section{Several statistical correlations for normal $\delta$ Scuti stars}

\begin{table*}
\footnotesize
\begin{center}
\caption{LAMOST observations of $\delta$ Scuti variable stars without periods listed in VSX.}\label{XXXX}
\begin{tabular}{lllllllll}\hline\hline
Name &$\alpha(2000)$ & $\delta(2000)$ &  Dates & Sp.  & $T$ (K)& $\log g$ &[Fe/H] & $V_{\rm r}$ (km s$^{-1}$) \\\hline
    GP Cnc                         & 129.5405 & 7.22583 &   2015-02-06  &         F0  & 7320  & 4.12  & -0.39  & -11 \\
    V0501 Per                      & 63.96912 & 51.21828 &   2012-02-08  &         F0  & 7140  & 4.26  & -0.14  & -13 \\
    UV Tri                         & 23.00046 & 30.36569 &   2013-10-04  &         F0  & 7250  & 3.89  & 0.14  & 25 \\
    NSV 1273                       & 56.39346 & 24.46328 &   2016-11-10  &       A6IV  & 7317  & 4.11  & -0.08  & 1.57 \\
    NSV 1273                       & 56.39346 & 24.46328 &   2016-09-30  &       A6IV  & 7294  & 4.08  & -0.10  & 2.94 \\
    $[EHV2002] 1-5-0832$           & 91.85333 & 45.9975 &   2013-12-15  &         F0  & 6980  & 4.11  & -0.39  & -41 \\
    HD 12899                       & 31.67329 & 14.58378 &   2012-09-30  &         F0  & 7170  & 3.85  & 0.06  & -15 \\
    HD 89503                       & 154.9601 & 7.61394 &   2016-01-22  &         F0  & 7540  & 3.79  & -0.08  & 1 \\
    HD 89503                       & 154.9601 & 7.61394 &   2015-04-24  &        A7V  & 7450  & 3.87  & -0.03  & -5 \\
    BD+19 572                      & 55.00588 & 19.81067 &   2016-01-29  &         K1  & 5070  & 2.84  & 0.05  & 17 \\
    BD+16 1693                     & 125.09 & 15.86461 &   2016-03-13  &         F0  & 7130  & 3.82  & 0.10  & 12 \\
    HD 221012                      & 352.119 & 18.69408 &   2014-11-05  &        A7V  & 7610  & 4.01  & 0.00  & 0 \\
\hline\hline
\end{tabular}
\end{center}
\end{table*}

In this section, we analyze the relationships between the pulsating period and those stellar atmospheric parameters (e.g., temperature, gravitational acceleration, and metallicity) and then investigate physical properties of the $\delta$ Scuti stars. Since the UCVs are quite different from the normal ones. When we investigate the physical properties of NDSTs, they should be excluded. As shown in Fig. 4, there is a good relation between the pulsating period ($P$) and the effective temperature ($T$) for NDSTs. A least-squares solution yields,
\begin{equation}
T=7233.5(\pm36.7)+1251.3(\pm242.0) \exp [-\frac{P}{0.033(\pm0.007)}].
\end{equation}
This relation (the solid red line) shows that the hotter $\delta$ Scuti stars tend to have extremely shorter periods than the late-type variables. It can be explained as an evolutionary effect because the hotter $\delta$ Scuti stars tend to be near the main sequence, while cooler variables are more evolved (e.g., Rodr\'{i}guez et al. 2000).

The correlation between the period and the gravitational acceleration $\log g$ is shown in Fig. 10 where 10 $\delta$ Scuti variable stars without period in VSX are not displayed. Their atmosphere parameters are listed in Table 7. Blue dots in the figure refer to NDSTs, while the green ones to UCVs. The two dashed lines are the borders of UCVs. It is well known that there is a basic relation for pulsating stars,
\begin{equation}
P\sqrt{\rho/{\rho}_{\odot}} = Q,
\end{equation}
where $\rho$ is the mean density, $Q$ is the pulsation constant. As stars evolving from zero-age main sequence, both the the mean density $\rho$ and the gravitational acceleration $\log g$ should be decreasing. Therefore, the pulsating period should be increasing according to equation (2). It is expected that there is a well correlation between the pulsating period and the gravitational acceleration, i.e., the long-period $\delta$ Scuti stars have low gravitational accelerations. As shown in Fig. 10, no such relation can be seen if all sample stars are considered. However, after UCVs are excluded, a well relation (the solid red line) between the period and the gravitational acceleration is detected for NDSTs with period shorter than 0.3\,d. A least-squares solution leads to the follow equation,
\begin{equation}
\log g = 4.046(\pm0.015)-1.168(\pm0.152)\times{P}.
\end{equation}
This relation is a strong observational evidence for the theoretically basic relation for pulsating stars.

\begin{figure}
\begin{center}
\includegraphics[width=1.15\columnwidth]{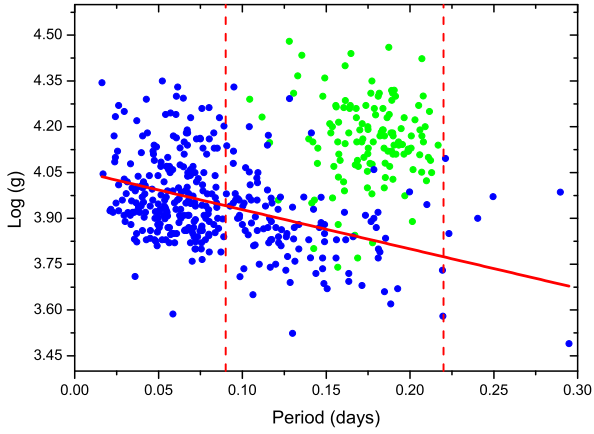}
\caption{The relation between the pulsating period and the gravitational acceleration for short-period NDSTs ($P<0.3$\,d). Blue dots refer of NDSTs, while green ones to UCVs that are between the two red dashed line. The red solid line represents the linear relation between the period and the gravitational acceleration. }
\end{center}
\end{figure}

\begin{figure}
\begin{center}
\includegraphics[width=1.15\columnwidth]{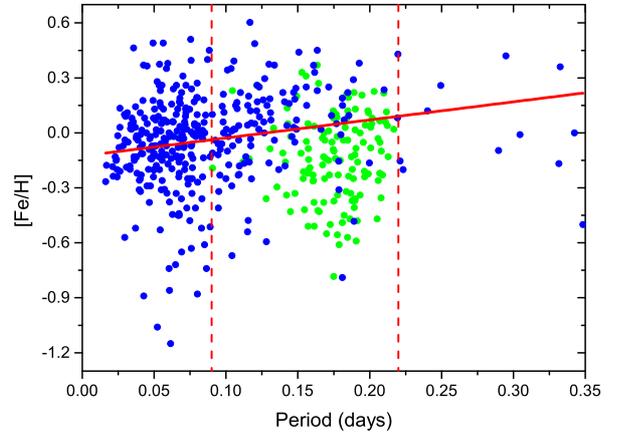}
\caption{The relation between the pulsating period and the metallicity [Fe/H] for short-period $\delta$ Scuti variable stars. Symbols are the same as those in Fig. 10. It is shown that there is a linear correlation between the pulsating period and the metallicity.}
\end{center}
\end{figure}

\begin{figure}
\begin{center}
\includegraphics[width=1.15\columnwidth]{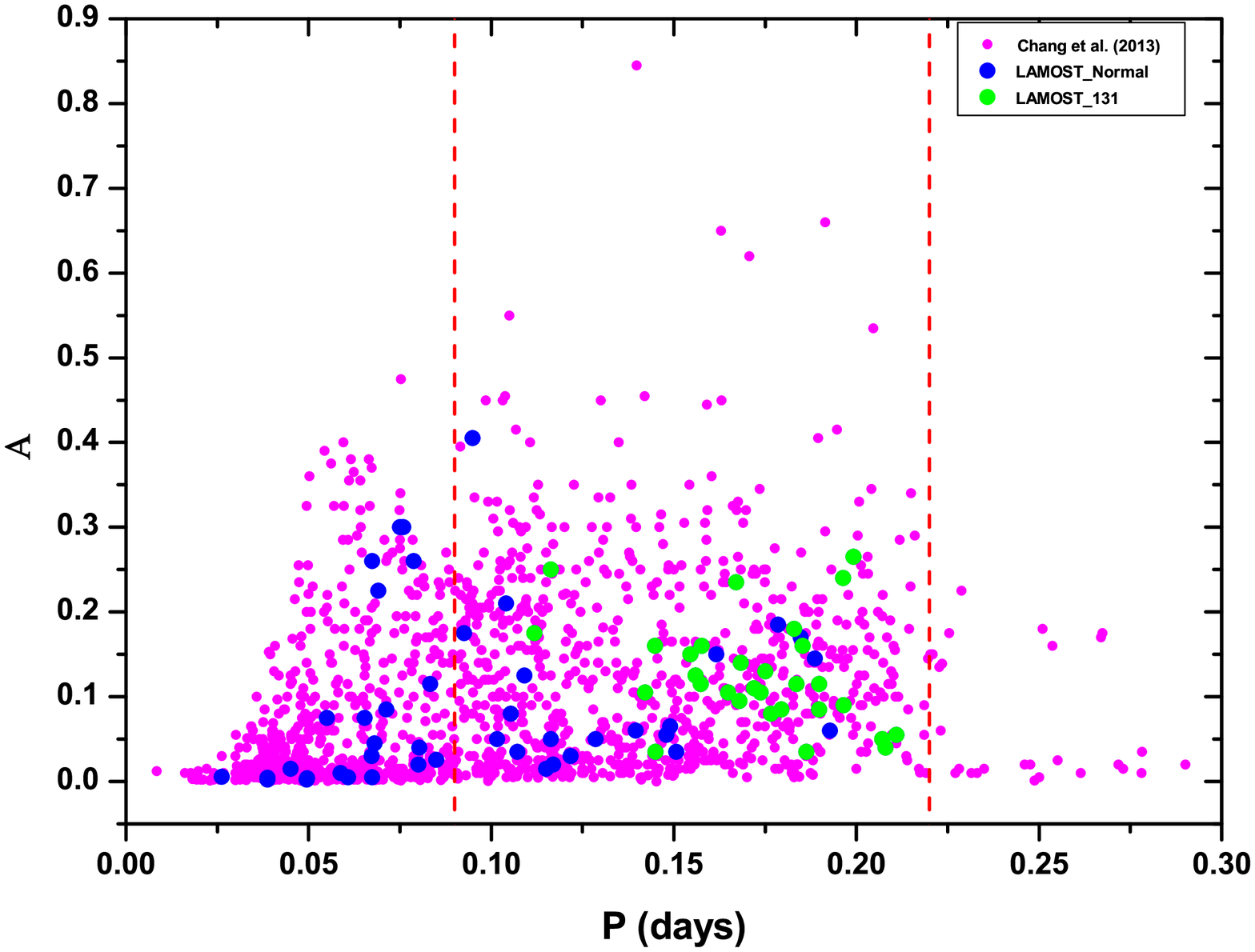}
\caption{The relation between the period and the photometric amplitude (A) for NDSTs and UCVs. Those V-band photometric amplitudes were from Chang et al. (2013). Blue dots refer of NDSTs, while green ones to UCVs. For comparison, the other targets are also shown as magenta dots.}
\end{center}
\end{figure}

The relation between the pulsating period and the metallicity [Fe/H] for short-period $\delta$ Scuti stars is plotted in Fig. 11. For comparison, those UCVs are also shown in the figure as green dots. For SX Phe-type pulsating stars in the globular clusters, there is a correlation between the metallicities and the periods of the variables (e.g., McNamara 1995, 1997; Rodr\'{i}guez \& Breger 2000). As shown in Fig. 11, the metallicity is weakly correlated with the period (the solid red line). By using the least-squares method, the following equation,
\begin{equation}
[\rm Fe/H]=-0.121(\pm0.026)+0.92(\pm0.25)\times{P},
\end{equation}
is derived after UCVs are excluded. The relation tell us that metal-poor stars enter the instability strip mostly with periods shorter than 0.1 d. By using equation (4), we could obtain [Fe/H]=0, when $P$=0.132\,d. This indicating that a $\delta$ Scuti star with a solar metallicity should has a typical period of P=0.132\,d. This statistical relation could be explained as that metal-poor pulsating stars evolving into pulsationally unstable states from main sequences displaced below the metal-strong main sequence (e.g., McNamara 1997).

Among the $\delta$ Scuti stars catalogued by Chang et al. (2013), 83 of them were observed by LAMOST including 33 UCVs and 50 NDSTs. By using the photometric data in the V-band collected by Chang et al. (2013), the correlation between the period and the photometric amplitude is displayed in Fig. 12 where blue dots refer of NDSTs, while green ones to UCVs. Apart from the variables observed by LAMOST, the other targets investigated by Chang et al. (2013) are also shown as magenta dots. As pointed out by Chang et al. (2013), the amplitudes are usually in the range of 0.003-0.9\,mag in the $V$ band and this figure tells us that there is no relation between amplitudes and periods for the field stars. Rodr\'{i}guez \& Breger (2001) mentioned that pulsating variables in the period range from 0.25 to 0.3\,d may need to be reclassified as evolved Population I $\delta$ Scuti or Population II RRc (or $\gamma$ Dor). As can be seen in Fig. 12, the amplitudes of UCVs are similar to those of NDSTs. The UCVs are solar-type cool variables that have periods in the range from 0.09 to 0.22\,d (the two dashed red lines). There are many $\delta$ Scuti stars whose pulsating periods are in the period range. It is possible that some of them belong to UCVs and they need to be reclassified.

\section{Discussions and conclusions}

\begin{figure}
\begin{center}
\includegraphics[width=1.15\columnwidth]{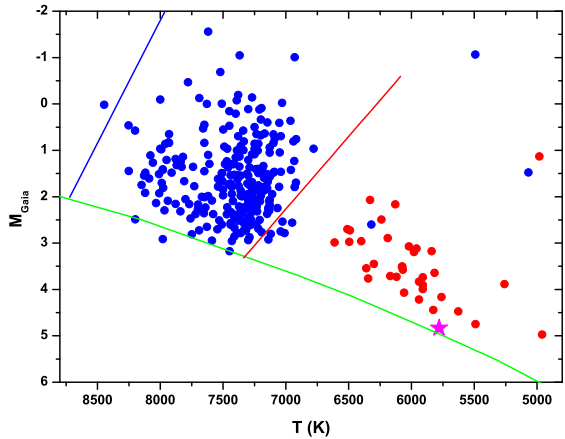}
\caption{The $H-R$ diagram for $\delta$ Scuti variable stars observed by both LAMOST and $GAIA$. Blue dots refer to NDSTs, while red dots to UCV ones. The position of the Sun is platted as the red star. The solid blue line and the red dashed line represent the blue and red edges of $\delta$ Scuti variables from McNamara (2000). The green line stands for zero-age main sequence from Kippenhahn et al. (2012).}
\end{center}
\end{figure}

\begin{figure}
\begin{center}
\includegraphics[width=1.15\columnwidth]{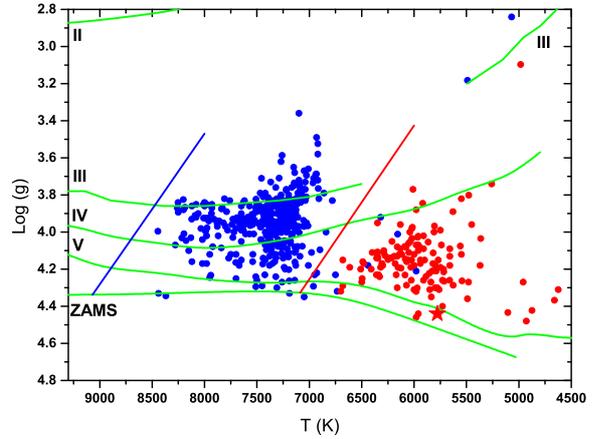}
\caption{The $\log g-T$ diagram for $\delta$ Scuti variable stars observed by LAMOST. Symbols are the same as those in Fig 12. The blue and red lines represent the blue and red edges of $\delta$ Scuti instability trip from Rodr\'{i}guez \& Breger (2001). The green lines stand for the luminosity classes range between II and V that are from Straizys \& Kuriliene (1981), while the zero-age main-sequence line is from Cox (2000).}
\end{center}
\end{figure}

In the past 20 years, the number of $\delta$ Scuti variable stars is enormously increasing. In addition to the immense effort developed by several photometric surveys on the ground, several space missions such as MOST (Walker et al. 2003), CoRoT (Baglin et al. 2006) and Kepler (Borucki et al. 2010) were carried out that led to the discovery of a lot of new variables and provide high-precision photometry. However, spectroscopic information for those Scuti stars are lack. Among 3689 $\delta$ Scuti variables listed in VSX catalogue, 766 of them were observed in LAMOST spectral survey from 2011 October 24 to 2017 June 16. We catalogue those $\delta$ Scuti stars and their spectral types are given. Stellar atmospheric parameters for 525 ones are presented. By analysing 32 $\delta$ Scuti variables observed four times or more, we show that the standard errors of the effective temperatures, the gravitational acceleration and the metallicity are usually lower than 100\,K, 0.1\,dex and 0.08\,dex, respectively. The atmospheric parameters for 352 stars were determined by Frasca et al. (2016) who mainly used high-resolution optical spectra collected from the literature. To check the LAMOST data (DR4 and the first three quarters of DR5) released recently, their LASP data are compared with those derived by Frasca et al. (2016). It is shown in Fig. 2 that there are very good agreements for the four stellar atmospheric parameters. The determined standard deviations are 135\,K for $T$, 0.21 dex for $\log g$, 0.14 dex for [Fe/H], and 11\,km/s for $V_{r}$, respectively. Meanwhile, the LASP data of the 188 $\delta$ Scuti stars in the LAMOST-Kepler field are also compared with those derived by using the ROTFIT code (Frasca et al. 2016). The results indicate that stellar atmospheric parameters for those $\delta$ Scuti stars derived by LAMOST are reliable.

As shown in Table 5, 50 $\delta$ Scuti stars have radial velocity differences larger than 15\,km s$^{-1}$ among 178 $\delta$ Scuti stars that were observed two times or more by LAMOST. They may be candidates of binary $\delta$ Scuti systems. Moreover, the radial velocities of some $\delta$ Scuti variables are very high. Table 6 shows that 38 $\delta$ Scuti variables have radial velocities higher than $V_{\rm r}=70$\,km s$^{-1}$. They may be also the members of binary or multiple systems. Some of those $\delta$ Scuti systems may be eclipsing binaries with high orbital inclinations. If they are real binary systems, the change of the relative distance of the $\delta$ Scuti variables from the Sun can result in the observed cyclic change in the O$-$C (the Observed$-$Computed maxima times) diagram when the $\delta$ Scuti stars orbit the barycenter of the binary system. Therefore, they could be confirmed by monitoring their maxima through the analyses of the light-travel time effect (e.g., Li et al. 2010, 2013; Qian et al. 2015).

By analysing stellar atmospheric parameters of $\delta$ Scuti pulsating stars derived by LAMOST, we show that there is a group of 131 UCVs with period in the range from 0.09 to 0.22\,d. The UCVs are solar-type stars with temperature lower than 6700\,K. It is found that there are two peaks on the distributions of the period, the effective temperature, the gravitational acceleration, and the the metallicity. One is the main peak, while the other is a small one. We show that the small peaks in those distributions are mainly caused by the existence of this group of variable stars. Those UCVs are more metal poor stars and have higher gravitational accelerations than those of NDSTs.

Among the 525 $\delta$ Scuti pulsating stars whose stellar atmospheric parameters were determined by LAMOST, 297 were also observed by $Gaia$ (Prusti et al. 2016) including 262 NDSTs and 35 UCVs. Their parallaxes and apparent magnitudes were given in Gaia Data Release 1 (Brown et al. 2016). The H-R diagram of the $\delta$ Scuti stars observed by both LAMOST and $Gaia$ is plotted in Fig. 13 where the effective temperature are from LAMOST, while the photometric absolute magnitudes are determined by using $Gaia$ data. The the blue and red edges for $\delta$ Scuti stars are from McNamara (2000). The green line in the figure represents the zero-age main sequence from Kippenhahn et al. 2012. As shown in the figure, apart from three targets, KIC\,9700679 ($P=3.81679$\,d), TAOS\,59.00115 ($P=0.038706$\,d), and $BD\,+19^{\circ}572$ (without period), the rest NDSTs locate inside the classical Cepheid instability strip, while the UCVs are far beyond the red edge of instability strip. It is shown that those UCVs are more evolved than the Sun.

The relations between the gravitational acceleration ($\log g$) and the effective temperature (T) for all variable stars observed by LAMOST is shown in Fig. 14 where blue dots refer to NDSTs, while red dots to UCVs. The position of the Sun is plotted as the red star. The blue and red lines in the figure stand for the borders of $\delta$ Scuti pulsating stars. Those green lines shown represent stellar luminosity classes range between II and V that are from Straizys \& Kuriliene (1981), while the zero-age main-sequence line is from Cox (2000). As that shown in the H-R diagram, it is found that all of the UCVs are beyond the red edge of $\delta$ Scuti instability trip. Most of them are evolved from zero-age main-sequence stars with luminosity classes range between IV and V. One UCV, KIC 8747415, may be giant or sub-giant with $\log g=3.097$.
There are several targets, e.g., KIC\,9700679 ($P=3.81679$\,d), KIC\,5446068 ($P=3.48432$\,d), BOKS-24178 ($P=0.7$\,d), ASAS\,J070907+2656.7 ($P=0.043322$\,d), TAOS\,59.00115 ($P=0.038706$\,d), and $BD\,+19^{\circ}572$ (without period), are beyond the red edge of $\delta$ Scuti stars with temperatures below 6700\,K. Their periods are not in the range between 0.09 and 0.22\,days. Although they were classified as NDSTs, their physical properties are special and need to be investigated in the future. As shown in Figs 13 and 14, there is a gap between zero-age main sequence and the position of NDSTs in the H-R diagram and in the $\log g-T$ diagram. The gaps increase with growing effective temperature (e.g., Balona \& Dziembowski 2011).

Those UCVs should be excluded when we investigate the physical properties of NDSTs because they are quite different from the others. It is found that there is a good relation between the effective temperature ($T$) and the pulsating period ($P$) for NDSTs. It shows that the temperature is rapidly decreasing when the period is increasing. This relation reveals that the hotter $\delta$ Scuti stars tend to be near the main sequence while cooler variables are more evolved (e.g., Rodr\'{i}guez et al. 2000). A well relation between the gravitational acceleration and the period is discovered for short-period $\delta$ Scuti stars with period shorter than 0.3\,d (solid red line in Fig. 10). The gravitational acceleration is decreasing as the period increasing. This relation is a direct observational evidence for the theoretically basic relation in equation (3) for pulsating stars. Moreover, the metallicity is detected to be weakly correlated with the period as that found for SX Phe-type pulsating stars in the globular clusters (e.g., McNamara 1997).

\begin{figure}
\begin{center}
\includegraphics[width=1.15\columnwidth]{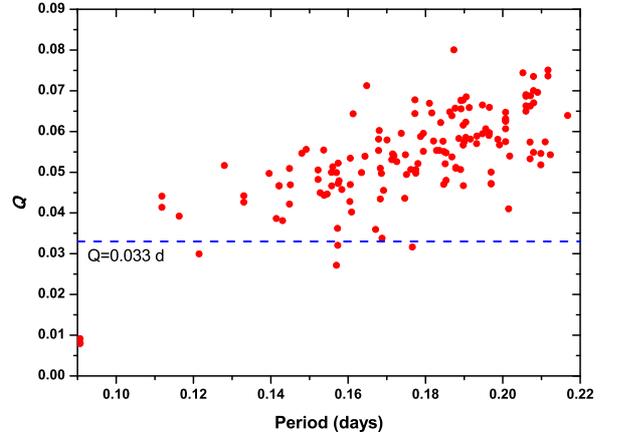}
\caption{The plot of the pulsation constant (Q) along with the period for the UCVs observed by LAMOST. The dashed red line refers to $Q=0.033$\,d. It is shown the values of Q for most of UCVs are larger than 0.033\,d indicating that they may not be $\delta$ Scuti pulsating stars.}
\end{center}
\end{figure}

\begin{figure}
\begin{center}
\includegraphics[width=1.15\columnwidth]{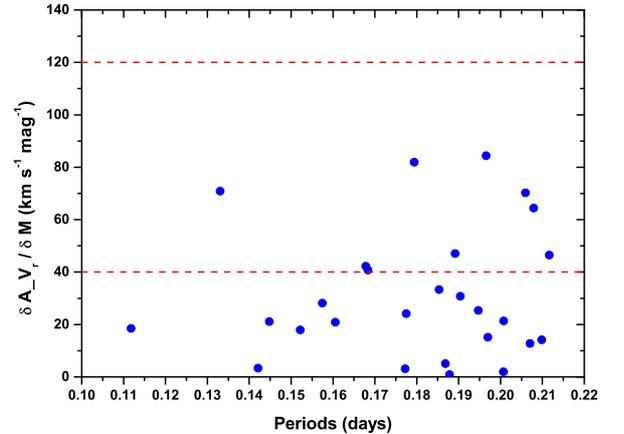}
\caption{A plot of the ratio of radial velocity to light amplitude along with the period for 28 UCVs observed two times or more. The two red dashed lines refer to the range for $\delta$ Scuti stars.}
\end{center}
\end{figure}

To check whether those UCVs are $\delta$ Scuti stars or not, their pulsation constants (Q) are calculated by using the following equation given by Breger (1990),
\begin{equation}
\log (Q/P) = 0.5 \log g +0.1 M_{bol} + \log T -6.456.
\end{equation}
The effective temperature $T$ and the gravitational acceleration $\log g$ are from the data of LAMOST, while $M_{bol}$ could be computed with the equation (e.g., McNamara 2011),
\begin{equation}
M_{bol} = -2.89 \log P -1.31.
\end{equation}
The results are shown in Fig. 15, where the dashed red line refers to $Q=0.033$\,d. As displayed in the figure, the values of $Q$ for most of UCVs are larger than 0.033\,days. As $\delta$ Scuti stars are p-mode and mixed-mode pulsators, $Q$ needs to be below 0.033\,d (e.g., Fitch 1981). Therefore, the derived results
may indicate that most UCVs do not belong to $\delta$ Scuti pulsating stars. The other method to check them is to determine the ratios of radial velocity to light amplitude of the stellar variability. If they are $\delta$ Scuti pulsating stars, the ratios should be in the range from 40 to 120 km$^{-1}$mag$^{-1}$ (e.g., Breger et al. 1976; Daszynska-Daszkiewicz et al. 2005). 28 UCVs were observed two times or more and the difference between the lowest and the highest radial velocities are determined. Then the ratios of radial velocity to light amplitude could be estimated and the results are plotted in Fig. 16. As shown in the figure, although some ratios are below 40 km$^{-1}$mag$^{-1}$, we could not rule out them as $\delta$ Scuti stars because our data are insufficient to derive the amplitude of the radial velocity curve.

Some UCVs are classified as both $\delta$ Scuti stars (DSCT) and EC-type contact binaries implying that they are actually suspected as contact binaries, i.e. their variability classification is not unique. Those suspected contact binary stars may have double periods, i.e., in the period range from 0.18 to 0.44\,d. This is typical range for EW-type contact binary systems (see fig. 1 in Qian et al. 2017a). Those binary stars are similar to contact systems investigated by Qian et al. (2007, 2013, 2014). Moreover, as listed in Table 7,
several UCVs show large radial velocity scatters indicating that they may be close binary stars. In this way, the derived effective temperatures for those sample stars are very useful during solving their light curves. The other atmospheric parameters (e.g., $\log g$ and [Fe/H]) will provide us valuable information to understand their formation and evolutionary state.
Our results show that about 25\% of the known $\delta$ Scuti pulsating stars may be misclassified. They may be close binary stars rather than $\delta$ Scuti stars. As shown in Fig. 1, about 1800 objects listed in VSX are classified as $\delta$ Scuti pulsating stars in the period range from 0.09 to 0.22\,d. Some of them may need to be reclassified.

The blue edge of the instability strip for $\delta$ Scuti stars is theoretically well constrained (e.g., Pamyatnykh 2000). However, the red edge is rather complicated and has a large range of possibilities for the slope and shape (e.g., Chang et al. 2013). The investigation of some authors (e.g., Houdek et al. 1999) indicated that oscillations in solar-like stars are intrinsically damped and stochastically driven by convection, while the calculations by Cheng \& Xiong (1997) using the theory of Xiong (1989) predicted over stable solar oscillations. If some UCVs are confirmed as pulsating stars, they may be a new-type pulsator. Their observations and investigations will shed light on theoretical instability domains and on the theories of interacting between the pulsation and the convection of solar-type stars. New observations and detailed investigations on UCVs are timely and required urgently in the near future.

\section*{Acknowledgements}

This work is partly supported by National Natural Science Foundation of China (No. 11325315). Guoshoujing Telescope (the Large Sky Area Multi-Object Fiber Spectroscopic Telescope LAMOST) is a National Major Scientific Project built by the Chinese Academy of Sciences. Funding for the project has been provided by the National Development and Reform Commission. LAMOST is operated and managed by the National Astronomical Observatories, Chinese Academy of Sciences. Spectroscopic observations used in the paper were obtained with
LAMOST from 2011 October 24 to 2016 November 30. This work has made use of data from the European Space Agency (ESA)
mission {\it Gaia} (\url{https://www.cosmos.esa.int/gaia}), processed by
the {\it Gaia} Data Processing and Analysis Consortium (DPAC;
\url{https://www.cosmos.esa.int/web/gaia/dpac/consortium}). Funding
for the DPAC has been provided by national institutions, in particular
the institutions participating in the {\it Gaia} Multilateral Agreement.

%%%%%%%%%%%%%%%%%%%%%%%%%%%%%%%%%%%%%%%%%%%%%%%%%%



%\bibliographystyle{mnras}


%%%%%%%%%%%%%%%%%%%%%%%%%%%%%%%%%%%%%%%%%%%%%%%%%%

%%%%%%%%%%%%%%%%% APPENDICES %%%%%%%%%%%%%%%%%%%%%

%\appendix



%%%%%%%%%%%%%%%%%%%%%%%%%%%%%%%%%%%%%%%%%%%%%%%%%%

% Don't change these lines
\bsp	% typesetting comment
\label{lastpage}
\end{document}